\documentclass[preprint,12pt]{elsarticle}

\usepackage{amssymb}
\usepackage{booktabs}
\usepackage{dcolumn}
\usepackage{multirow}
\usepackage{adjustbox}
\usepackage{subfig}
\usepackage{graphicx}
\usepackage{natbib} 
\usepackage[obeyspaces,hyphens]{url}
\usepackage{booktabs}
\usepackage{lscape}
\usepackage{nicefrac}
\usepackage{xcolor}
\usepackage[countmax]{subfloat}

\biboptions{sort&compress}

\newcommand{\be}{\begin{equation}}
\newcommand{\ee}{\end{equation}}
\newcommand{\bse}{\begin{subequations}}
\newcommand{\ese}{\end{subequations}}
\newcommand{\beq}{\begin{eqnarray}}
\newcommand{\eeq}{\end{eqnarray}}
\newcommand{\lb}[1]{\label{#1}}

\newcommand{\dn}[1]{\raisebox{-0.7mm}{$\scriptstyle #1 $}}

\newcommand{\dd}{{\mathrm{d}}}
\usepackage{amsmath}
\usepackage{txfonts}
\usepackage{epsfig}
\usepackage{graphicx}
\usepackage{empheq}
\journal{}

\begin{document}

\begin{frontmatter}

\title{Modeling Income Distribution with the Gause-Witt Population Ecology System}
\author[1]{Marcelo B.\ Ribeiro\corref{cor3}%
}
\address[1]{Physics Institute, Universidade Federal do Rio de Janeiro,
            Rio de Janeiro, Brazil}
\ead{mbr@if.ufrj.br}
\cortext[cor3]{Orcid 0000-0002-6919-2624}

\begin{abstract}
This paper presents an empirical application of the
Gause-Witt model of population ecology and ecosystems to the income
distribution competitive dynamics of social classes in economic
systems. The Gause-Witt mathematical system of coupled nonlinear
first-order ordinary differential equations employed to model
population of species sharing the same ecological niche and
competing for the same resources was applied to the income data
of Brazil. Previous studies using Brazilian income data from 1981
to 2009 showed that the complementary cumulative distribution
functions built from yearly datasets have two distinct segments:
the lower income region comprising of about 99\% of the population
can be represented by the Gompertz curve, whereas the richest 1\%
is described by the Pareto power-law. The results of applying the
Gause-Witt system to Brazilian income data in order to describe
the distributive competition dynamics of these two population
shares indicate that the 99\% and 1\% income classes are mostly
in the dynamic state of stable coexistence.
\end{abstract}

\end{frontmatter}


\section{Introduction}

The growing perception that the gap between rich and poor has been
widening worldwide at galloping pace as a result of the so-called
globalization processes occurring in the last two to three decades
has raised concerns in many countries regarding the fairness of
income and wealth distribution. This has in fact become a hotly
disputed topic \cite{oxfam3,oxfam2} because the existence or not of
economic fairness in modern economies is a subject that goes to the
heart of any society's viewpoints on issues regarding social
opportunity and egalitarianism. This concern inevitably raised again
the interest of scholars who then started paying renewed attention
to this subject, and as result these issues became important research
themes among economists \cite{piketty} and econophysicists
\cite{incomedistro,ribeiro2020income,ghosh2023,greenberg2024}. Income
distribution has particularly been scrutinized with studies focusing
on its features, dynamic behavior and evolution \cite{ribeiro2020income}.

Studies characterizing the income distribution are mostly carried out
by means of the \textit{complementary cumulative distribution function}
(CCDF), a very useful way of data visualization because it provides
especial insights on the data's tail behavior \cite{mexico}. Various
works on the complementary cumulative income distribution are available
in the literature, and these analyses showed a clear pattern constituted
by the existence of two income segments in apparently most, if not all,
societies \cite[see][Sec.\ 2.4, and references therein]{ribeiro2020income}.
The first one comprises about 99\% of the lower income population, whereas
the second segment includes the remaining 1\% that forms the rich group.
So, it seems that as far as income is concerned most modern societies
are, broadly speaking, actually divided in two clearly distinct social
classes, usually called ``the 99\%'' and ``the 1\%'' \cite{oxfam1}.
Several studies indicate that the richest 1\% stratum characterized
by the CCDF tail is well represented analytically by a power-law as
the Italian civil engineer-turned-economist Vilfredo Pareto (1848-1923)
argued over a century ago \cite[][p.\ 312]{pareto1964cours}. The
remaining CCDF segment due to the 99\% populational portion can be
represented by other functions, the exponential and the Gompertz curve
\cite{gompertz,winsor} being two of the simplest among them
\cite[][Secs.\ 2.3, 2.4]{ribeiro2020income}.

Most studies on the 99-1\% income divide were basically concerned with
the functional CCDF characterization, a focus also present in Refs.\
\cite{nm09,fnm2010} which analyzed the income distribution of Brazil
and concluded that the 99\% segment can be well represented by the
Gompertz curve, whereas the 1\% rich portion at the CCDF's tail is
functionally described by the Pareto power-law. Afterwards, Ref.\
\cite{nm2013} attempted to model the Brazilian income distribution as a
competitive \textit{predator-prey system} to see if this model could
shed light in the evolving dynamics of this segmentation. Basically,
the variables of the system were defined as being on one hand the labor
share of the total, the Gompertzian 99\%, and employment rate on the
other, and these variables were noticed as tending to oscillate in an
almost limit cycle fashion in the phase space \cite{nm2013}, a behavior
that suggested a predator-prey like dynamics \cite[][Chap.\ 7]{kot}.

The predator-prey system when applied to economics became known as the
\textit{Goodwin model} \cite[Sec.\ 5.3]{ribeiro2020income} \cite{g67}
\cite[Sec.\ 24.4.3]{gandolfo} and there already is a fair amount of
empirical studies on the economic applicability of this model \cite[]
[Sec.\ 5.4]{ribeiro2020income}. Nevertheless, most studies that actually
try to test this model start with econometric quantities, that is,
considering the original parametric definitions as derived from quantities
like wage share, profit level, labor force, investment return, etc
\cite[see][Sec.\ 2, for a review of the literature]{nezami}. Moura Jr.\
\& Ribeiro \cite{nm2013} took a different testing route when they
identified the Gompertzian 99\% as one of the variables of the
predator-prey system. Despite this novelty, their results were in line
with previous testings, that is, the model showed qualitative support
but failed quantitatively \cite[][Secs.\ 5.4, 5.5]{ribeiro2020income}.

The present paper attempts to go a step further than Moura Jr.\ \& Ribeiro
\cite{nm2013}. The motivation is similar to the Goodwin model, which
borrowed the predator-prey system from population ecology in order
to model some economic features. The basic idea here is similar in the
sense that it attempts to use a population ecology model, but instead
of simple competition of one species feeding on the other this study
goes to the next modeling generalization level, that is, it considers
species living together in a certain ecosystem and consuming its resources.
In other words, the aim is to model the income distribution using the
ecological concept of resource ecosystem competition. Therefore, the
99-1\% income divide will be viewed as two income classes in competition
for the same economic resources. To do so the \textit{Gause-Witt model}
of species competition will be used with the data already provided by
Ref.\ \cite{nm2013}. The main result coming out of this analysis in that
in the period from 1982 to 2008, the time span of the available data,
the 99-1\% income divide is mostly in the dynamic state of \textit{stable
coexistence} as far as Brazil is concerned. This study indicates that the
modeling presented here can be applied to similar class divides of other
countries and regions, and at different time spans, as well as to economic
sectors in competition for a share of the same market.

The plan of the paper is as follows. Section \ref{eco} presents a brief
review of some basic concepts of mathematical ecology of populations
required in the present study. Section \ref{br} applies the Gause-Witt
model to Brazilian income data and Section \ref{discu} discusses the
results, their possible consequences to income distribution dynamics
and, more broadly, social class dynamics in economics. Sect.\
\ref{conc} presents the conclusions of this study, some caveats,
a few related literature, and possible future directions.

\section{Brief Aspects of Mathematical Ecology of Populations}\lb{eco}

Let us start with the following nonlinear system of coupled first-order
ordinary differential equations (ODEs):
\begin{subequations}
\begin{empheq}[left=\empheqlbrace]{align}
\displaystyle\frac{\dot{u}}{u}&=f_{1}(u,v), \lb{f1} \\[-8pt] \nonumber \\
\displaystyle\frac{\dot{v}}{v}&=f_{2}(u,v), \lb{f2}
\end{empheq}
\end{subequations}
where $\cdot \equiv \dd/\dd t$. This is written with a ratio on the
left hand side for simplicity and clarity in what follows. Specific
choices of the functions $f_1$ and $f_2$ of the system above have been
used in population ecology to model the dynamic interaction of species
in ecosystems for over a century.

\subsection{Lotka-Volterra Predator-Prey Model}\lb{lv}

The simplest choice such that the system remains coupled is the linear
approximation of both unknown functions. Hence, the expressions above
can be written as follows,
\bse
\begin{empheq}[left=\empheqlbrace]{align}
\displaystyle\frac{\dot{u}}{u}&=a_1-b_1v, \lb{f1lv} \\[-8pt] \nonumber \\
\displaystyle\frac{\dot{v}}{v}&=a_2-b_2u, \lb{f2lv}
\end{empheq}
\ese
where $a_1$, $b_1$, $a_2$ and $b_2$ are positive parameters. This ODE
system forms the well-known \textit{Lotka-Volterra model} of predator
and prey competition, which captures the empirical fact that the
population number of species in predator-prey systems represented by the
variables $u$ and $v$ have a tendency to oscillate. Its dynamic details
and empirical ecological support are comprehensively exposed in various
textbooks of mathematical ecology
\cite[Chap.\ 7]{kot}
\cite[Chap.\ 13]{case}
\cite[Chap.\ 6]{vande}
\cite[Secs.\ 6.2, 8.7]{edel}
\cite[Chap.\ 8]{pastor}
\cite[Chap.\ 10]{rockwood2}.

The Lotka-Volterra model was originally proposed to represent ecological
competition between predators feeding on preys. Despite this it found its
way into other areas where quantities are observed to oscillate having
limit cycles in the variables' phase space. This is, for instance, the
case of historical dynamics \cite[Sec.\ 2.1.3, p.\ 153]{turchin}, economics
\cite{g67} and econophysics \cite{nm2013,turcos}. In economics and
econophysics the predator-prey system became known as the
\textit{Goodwin growth-cycle model} \cite[Secs.\ 5.3, 5.4]{ribeiro2020income}
\cite{nm2013} \cite[Sec.\ 24.4.3]{gandolfo}.

\subsection{Gause-Witt Model}\lb{gw}

Before we discuss a second possible generalization of the system formed
by Eqs.\ (\ref{f1}) and (\ref{f2}) it is necessary first a little
digression about an important concept that will be required in the next
sections.

The Russian biologist Georgy Frantsevich Gause (1910-1986) carried out
a series of laboratory experiments on the growth of species sharing
the same ecological niche and competing for the same food, whose
results were published in a monograph entitled \textit{The Struggle
for Existence} \cite{gause34}. On the basis of his findings one can
reach at the following conclusions \cite[pp.\ 179-180]{rockwood2}
\cite[Sec.\ 9.1]{royama}.
\begin{enumerate}[(1)]
 \item Two species cannot coexist unless they are doing things
       differently.
 \item No two species can occupy the same ecological niche.
\end{enumerate}
These empirical findings form the core of the \textit{competitive
exclusion principle} \cite{hardin}, or \textit{Gause's principle
(law)}, fundamental in ecological thinking. It can be stated as
follows.
\begin{enumerate}[(a)]
 \item No two species of similar ecology can coexist or occupy the
       same ecological niche.
 \item Species which are complete competitors, that is, whose niches
       overlap completely, cannot coexist indefinitely.
\end{enumerate}
These two propositions basically state the same concept in slightly
different ways. Competitive exclusion plays a major role in evolution
because ``competition between species is an ecological process with
evolutionary consequences'' \cite[p.\ 33]{peter}.

The digression above shows the importance of the second, possibly
simplest, choices for the functions $f_1$ and $f_2$ as discussed by
Gause \cite[p.\ 47]{gause34},
\bse
\begin{empheq}[left=\empheqlbrace]{align}
\displaystyle\frac{\dot{u}}{u}&=a_1-b_1v-c_1u, \lb{f1gw} \\[-8pt] \nonumber \\
\displaystyle\frac{\dot{v}}{v}&=a_2-b_2u-c_2v, \lb{f2gw}
\end{empheq}
\ese
where $c_1$ and $c_2$ are also positive parameters. These expressions
had their dynamic properties analyzed by Gause in collaboration with the
Russian mathematical physicist Alexander Adolfowitsch Witt (1902-1938)
by constructing their respective $u-v$ phase spaces. The system of
nonlinear ODEs formed by Eqs.\ (\ref{f1gw})--(\ref{f2gw}) is called
the \textit{Gause-Witt model} \citetext{\citealp{gause-witt}; see also
\citealp[Sec.\ 2.1]{peter}}.

The importance of this model comes from the fact that differently from
the predator-prey system it does no depict one species feeding on the 
other, but competition of two species for resources available in the
same ecological niche, that is, the same ecosystem. So, it goes a step
further than the predator-prey model. In addition, it intrinsically
includes the concept of natural selection because Gause's experimental
work was focused on ``the mechanism of Darwinian natural selection
rather than merely about ecological competition (\ldots) His isocline
graphical method for analyzing competitive equilibria (\ldots) was
explicitly titled as a paper on natural selection'' \cite[][p.\ 631]
{mallet}. This property of resource competition and its associated
principle of competitive exclusion are the factors suggesting its
usefulness in modeling the process of income distribution competition
as we shall see in the next sections.
 
Although Volterra \cite{volterra26} and Lotka \cite{lotka32} discussed
the problem of species competition, the isoclines phase portrait of the
model (see Sec.\ \ref{phasegw} below) was more clearly exposed by Gause
and Witt in Ref.\ \cite{gause-witt}. In addition, since Gause provided a
clear empirical support for species competition in his seminal monograph
\cite{gause34}, following Ref.\ \cite[][Sec.\ 2.1]{peter} it seems
appropriate to call the system formed by Eqs.\ (\ref{f1gw}) and
(\ref{f2gw}) as the Gause-Witt model. Nevertheless, attributions for this
model are only made to Lokta and Volterra in several ecology textbooks,
with Ref.\ \cite{gause-witt} usually nowhere to be found in many texts.
For instance, Rockwood \cite[Sec.\ 7.3]{rockwood2} and several other
mathematical biology and ecology textbooks call the Gause-Witt model as
being a generalized Lotka-Volterra model. Kot \cite[pp.\ 198-203]{kot}
and Vandermeer \& Goldberg \cite[p.\ 225]{vande} both call it as
``Lotka-Volterra competition equations.'' Edelstein-Keshet \cite{edel}
calls it ``Lotka-Volterra model for species competition,'' and Pastor
\cite[Chap.\ 9]{pastor} calls it ``generalized Lotka-Volterra'' without
citing both Gause and Witt.

\subsection{Hutchinson Model}\lb{hutch}

The British ecologist George Evelyn Hutchinson (1903-1991) went a step
further than the model above and proposed the following system \cite{hut},
\bse
\begin{empheq}[left=\empheqlbrace]{align}
\displaystyle\frac{\dot{u}}{u}&=a_1-b_1v^2-c_1u, \lb{f1h} \\[-8pt] \nonumber \\
\displaystyle\frac{\dot{v}}{v}&=a_2-b_2u^2-c_2v. \lb{f2h}
\end{empheq}
\ese
The Lotka-Volterra model discussed predation. The Gause-Witt model
introduced the concept of resource competition. Hence, the question
posed by Hutchinson was the possibility that two species could also
exhibit higher order social interactions, that is, other types of
symbiotic relationships rather than predation and competition, such
as mutualism, commensalism, amensalism, neutralism, parasitism and
herbivory \cite[][p.\ 233]{princetonguide}. Basically the intensity
of the social interaction depends on the population densities
\cite[Sec.\ 2.2]{peter}. Nevertheless, objections were raised of whether
or not the \textit{Hutchinson model} given by Eqs.\ (\ref{f1h}) and
(\ref{f2h}) provides what it intends, and the resulting revision of
the model led to an even higher degree of nonlinearity \cite[Sec.\ 2.2]
{peter}.

Because the Hutchinson model is mathematically more complex, it is being
briefly presented here to show that there still are avenues to be
explored for higher order effects, explorations which apparently will
entail the use of more advanced mathematical tools \cite{peter,rutz2}.
Clearly Eqs.\ (\ref{f1}) and (\ref{f2}) can have further generalizations
beyond those presented here, although interpretation and application
become less obvious with increasing complexity and nonlinearity. Hence, 
moving from Lotka-Volterra to Gause-Witt will suffice for the purposes of
this paper, which are application and data testing.

\subsection{Phase Portrait Dynamics of the Gause-Witt model}\lb{phasegw}

The system formed by Eqs.\ (\ref{f1gw}) and (\ref{f2gw}) does not
have a trivial integration. Nevertheless, several of its dynamic features
can be revealed by means of the isocline graphical method introduced by
Gause and Witt \cite{gause-witt}. Let us briefly review it below.

Assuming $u$ and $v$ to be the population number or density of each
competing species in a Gause-Witt system, an ``equilibrium'' analysis
can be done when competition is complete and both species no longer grow.
Then the following results hold, 
\begin{subequations}
\begin{empheq}[left=\empheqlbrace]{align}
\dot{u}=0 \; \Longrightarrow \; a_1-b_1v-c_1u&=0, \; \iff \;
	{\dd v}/{\dd u}=\infty \;\; (i.e. \;\; {\dd u}/{\dd v}=0),
	\lb{equi1}\\[-12pt]\nonumber\\
\dot{v}=0 \; \Longrightarrow \; a_2-b_2u-c_2v&=0, \; \iff \;
	{\dd v}/{\dd u}=0, \lb{equi2}
\end{empheq}
\end{subequations}
since the phase plane is by definition written as below,
\be
\frac{\dd v}{\dd u}=\frac{v\,(a_2-b_2u-c_2v)}{u\,(a_1-b_1v-c_1u)}.
\lb{dvdu}
\ee
Eq.\ (\ref{equi1}) is called the \textit{vertical isocline} and Eq.\
(\ref{equi2}) the \textit{horizontal isocline}. The expression
(\ref{equi1}) leads to the following results, 
\bse
\begin{empheq}[left=\empheqlbrace]{align}
u=0 \; \Longrightarrow \; v&={a_1}/{b_1},\lb{equi3} \\[-12pt]\nonumber\\
v=0 \; \Longrightarrow \; u&={a_1}/{c_1},\lb{equi4}
\end{empheq}
\ese
whereas Eq.\ (\ref{equi2}) yields,
\bse
\begin{empheq}[left=\empheqlbrace]{align}
u=0 \; \Longrightarrow \; v&={a_2}/{c_2},\lb{equi5} \\[-12pt]\nonumber\\
v=0 \; \Longrightarrow \; u&={a_2}/{b_2}.\lb{equi6}
\end{empheq}
\ese

These results can be put graphically in the system's phase portrait.
Isoclines are curves where all zero equilibria of $u$ and $v$ occur,
but which are not, in general, solution curves of the system. Actually, 
solution curves cross vertical isoclines vertically and horizontal
isoclines horizontally.

Fig.\ \ref{blue-red} \textit{left} shows the \textit{nullcline} of
population number or density of species $v$, that is, the straight
line along which this species numbers do not go up or down as one
moves along the blue line in the phase plane. Horizontal isoclines mean
$\dd v/\dd u=0$ in the left plot. Fig.\ \ref{blue-red} \textit{right}
shows similar situation for species $u$, whose vertical isoclines are
given by $\dd u/\dd v=0$ and the respective nullcline for zero growth
of this species is shown as a red straight line.
\begin{figure}[ht]
\begin{center}$
\begin{array}{cc}
\includegraphics[scale=0.275]{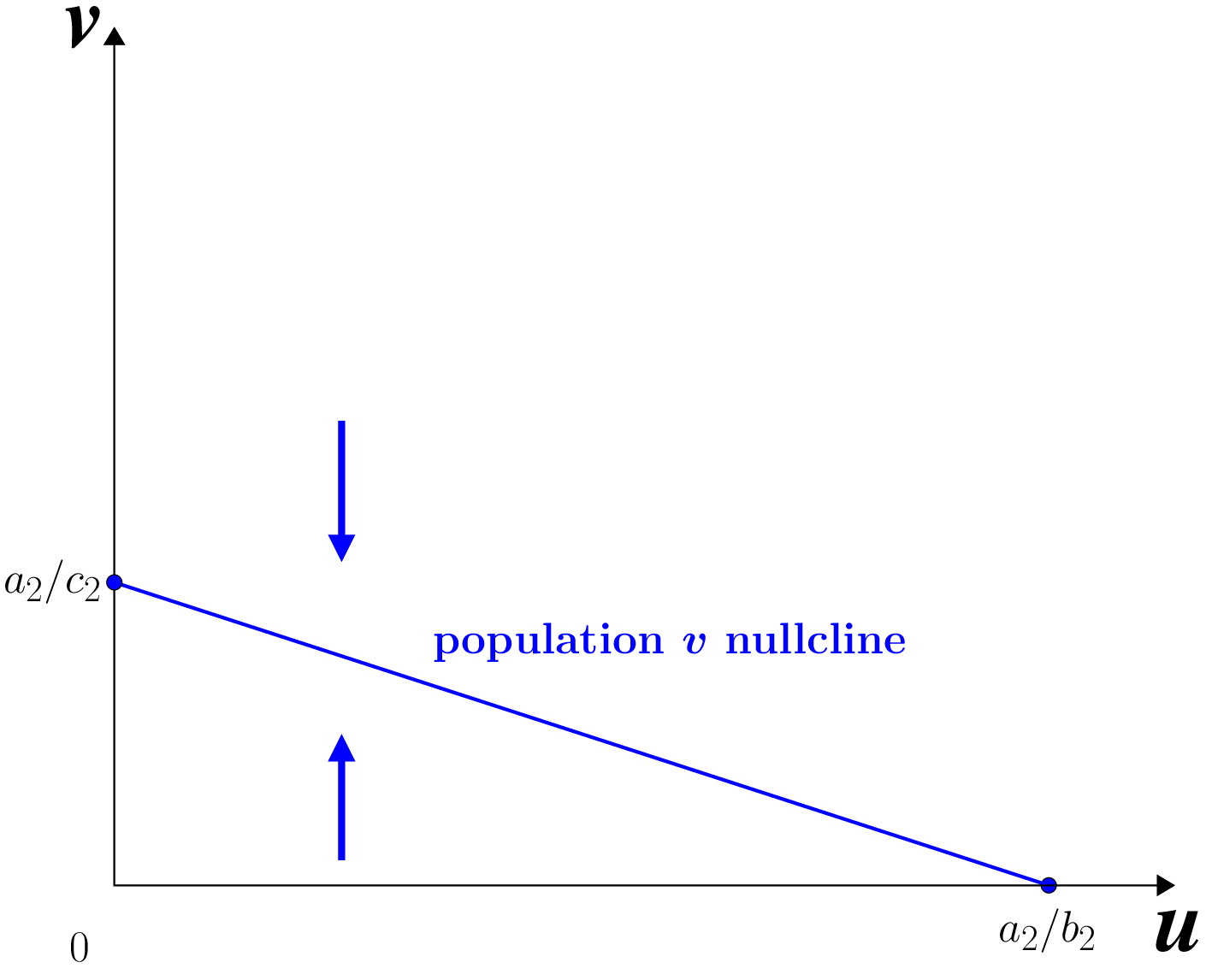} &
\includegraphics[scale=0.275]{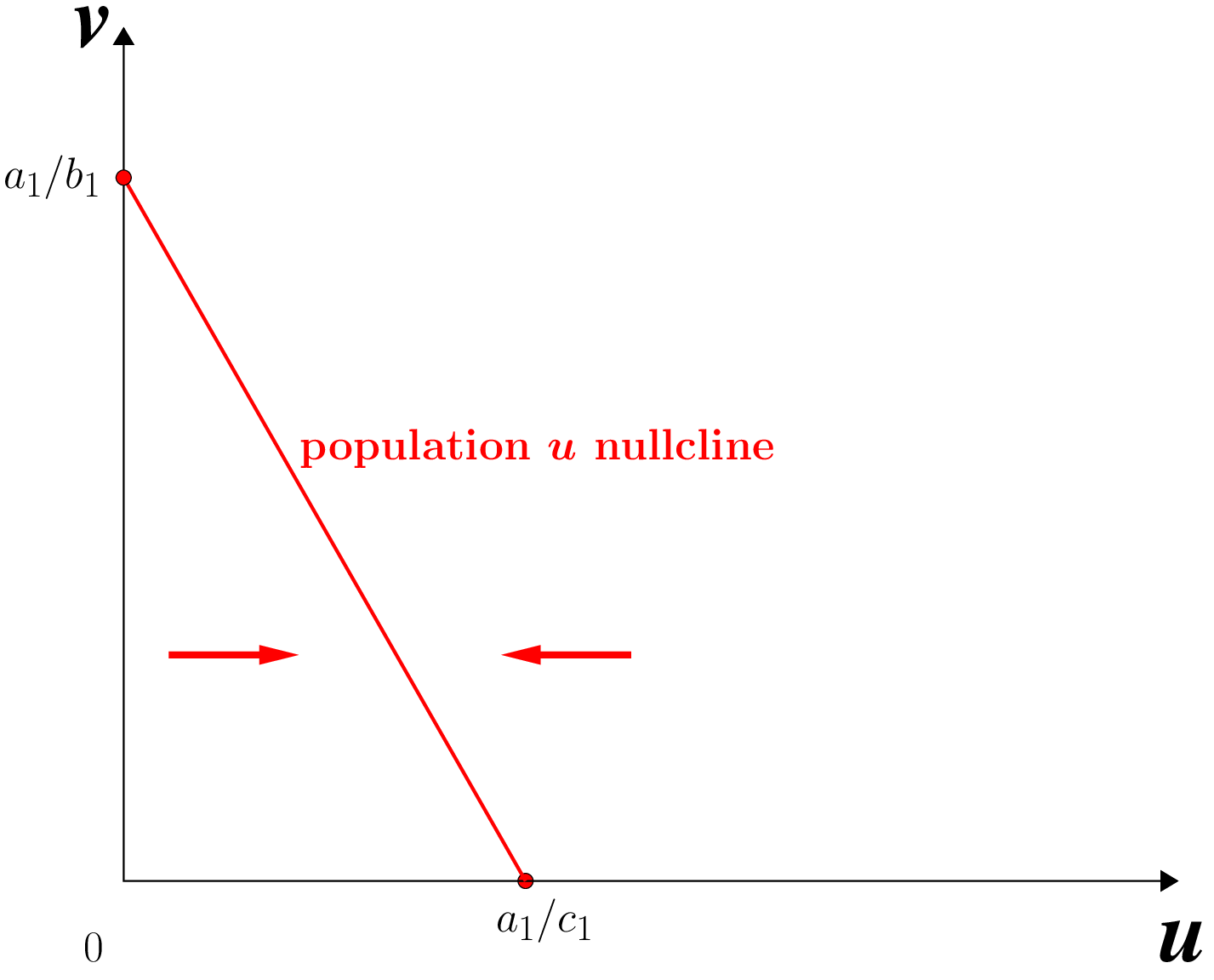} 
\end{array}$
\end{center}
\caption{\textit{Left}: The blue line is the zero isocline (nullcline)
    of population $v$, defined by $\dot{v}=0$ (Eqs.\ \ref{equi2},
    \ref{equi5}, \ref{equi6}). This isocline, defined by the straight
    line $\bigl[ v=a_2/c_2-(b_2/c_2)u \bigr]$, represents the saturation
    level of population $v$. The arrow on the lower side of the
    isocline represents a vector field which means that population $v$
    has enough resources to grow upwards, whereas the upper arrow means
    the opposite, \textit{i.e.}, that population $v$ has grown beyond
    the resources available to sustain it and has to decrease until it
    reaches the saturation level given by the isocline. \textit{Right}:
    The red line defines the zero isocline (nullcline) of population $u$,
    defined by the straight line equation $\bigl[ v=a_1/b_1-(c_1/b_1)u
    \bigr]$ when $\dot{u}=0$ (Eqs.\ \ref{equi1}, \ref{equi3},
    \ref{equi4}) where population $u$ reaches its saturation level. The
    left arrow indicates that this population can grow up to the
    nullcline, whereas the right arrow means that population $u$ no
    longer has enough resources to sustain it and has to decrease until
    it reaches the saturation level given by the nullcline.}
\lb{blue-red}
\end{figure}

Now considering the coordinates given by the resulting expressions
(\ref{equi3}) to (\ref{equi6}) the two plots of Fig.\ \ref{blue-red}
can be combined into four possible situations, as follows: 
\bse
\begin{empheq}[left=\empheqlbrace]{align}
a_1/b_1>a_2/c_2 \;\; \mathrm{and} \;\; a_1/c_1<a_2/b_2; \lb{stable}
\\[-12pt]\nonumber\\
a_1/b_1<a_2/c_2 \;\; \mathrm{and} \;\; a_1/c_1>a_2/b_2; \lb{unstable}
\\[-12pt]\nonumber\\
a_1/b_1<a_2/c_2 \;\; \mathrm{and} \;\; a_1/c_1<a_2/b_2. \lb{exclusaou} 
\\[-12pt]\nonumber\\
a_1/b_1>a_2/c_2 \;\; \mathrm{and} \;\; a_1/c_1>a_2/b_2; \lb{exclusaov}
\end{empheq}
\ese
Fig.\ \ref{stable-unstable} shows the first two combinations: nullclines,
or species saturation levels, for the expressions (\ref{stable}) and
(\ref{unstable}), respectively left and right plots. The \textit{left}
plot indicates that as the species represented by the variable $u$
approaches its respective nullcline horizontally, and the species given
by the variable $v$ does that vertically, the resultant dynamic, shown
in black arrows, leads to an endpoint of stable coexistence. The
\textit{right} plot of Fig.\ \ref{stable-unstable} shows that there is a
critical line of stability such that if any of the species goes further
away from it the dynamics leads to the elimination of one or the other.
\begin{figure}[ht]
\begin{center}$
\begin{array}{cc}
\includegraphics[scale=0.27]{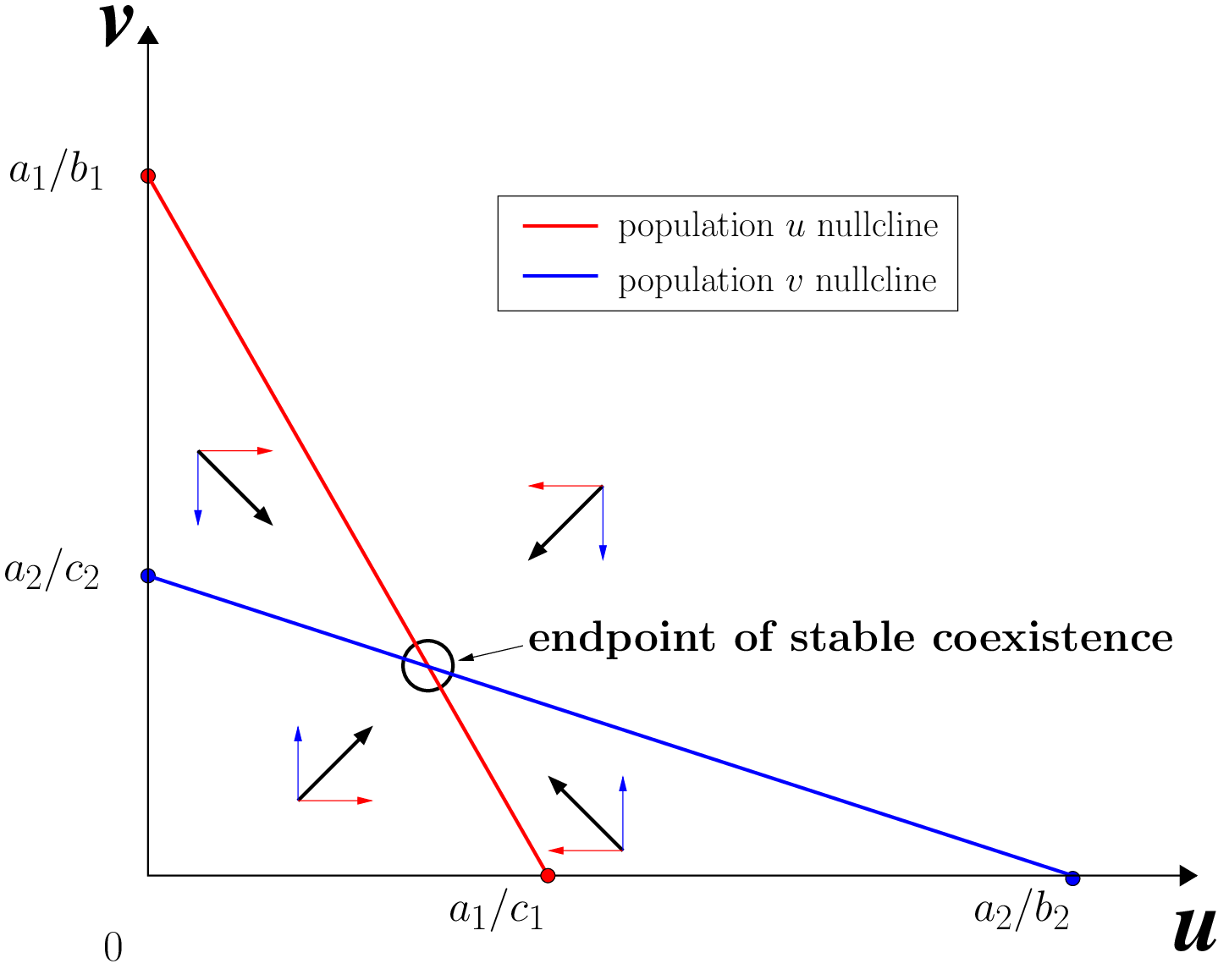} &
\includegraphics[scale=0.27]{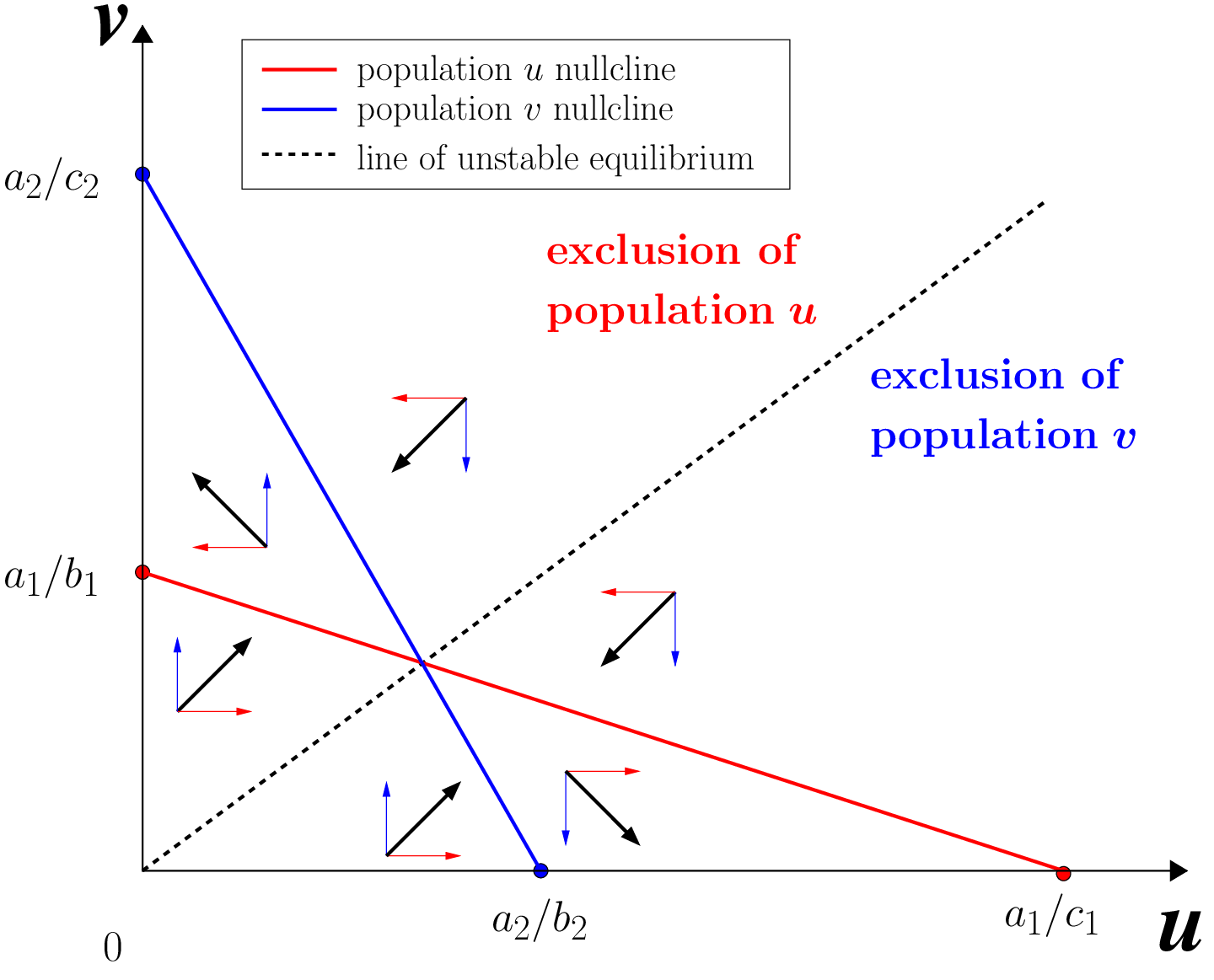} 
\end{array}$
\end{center}
\caption{\textit{Left}: Stable coexistence (inequalities \ref{stable}).
    The vector field of species $u$ and $v$ are respectively the red
    and blue arrows that correspond to the increase or decrease of
    each population depending on their positions in the phase plane.
    Black arrows are the resultants of the vector field at each region
    of the phase portrait. In this case both species end up coexisting
    because each species slows its own growth more than that of its
    competitor and the system converges to the end point of stable
    coexistence. \textit{Right}: Unstable coexistence (inequalities
    \ref{unstable}). Arrow colors mean the same as in the previous case.
    However, the equilibrium line is unstable because outside of it, 
    and  depending on where the system is located in the $u-v$ phase
    plane, one species eliminates the other by consuming all resources
    until the other species starves to death and is then competitively
    excluded from the ecosystem.}
\lb{stable-unstable}
\end{figure}

The other two remaining possible cases as given by the inequalities
(\ref{exclusaou}) and (\ref{exclusaov}) are shown in Fig.\
\ref{exclusions}, respectively left and right. The dynamic case
represented by the \textit{left} graph leads to exclusion of the
population $u$, whereas the \textit{right} plot results in the
extinction of the population $v$ by competitive exclusion.
\begin{figure}[ht]
\begin{center}$
\begin{array}{cc}
\includegraphics[scale=0.27]{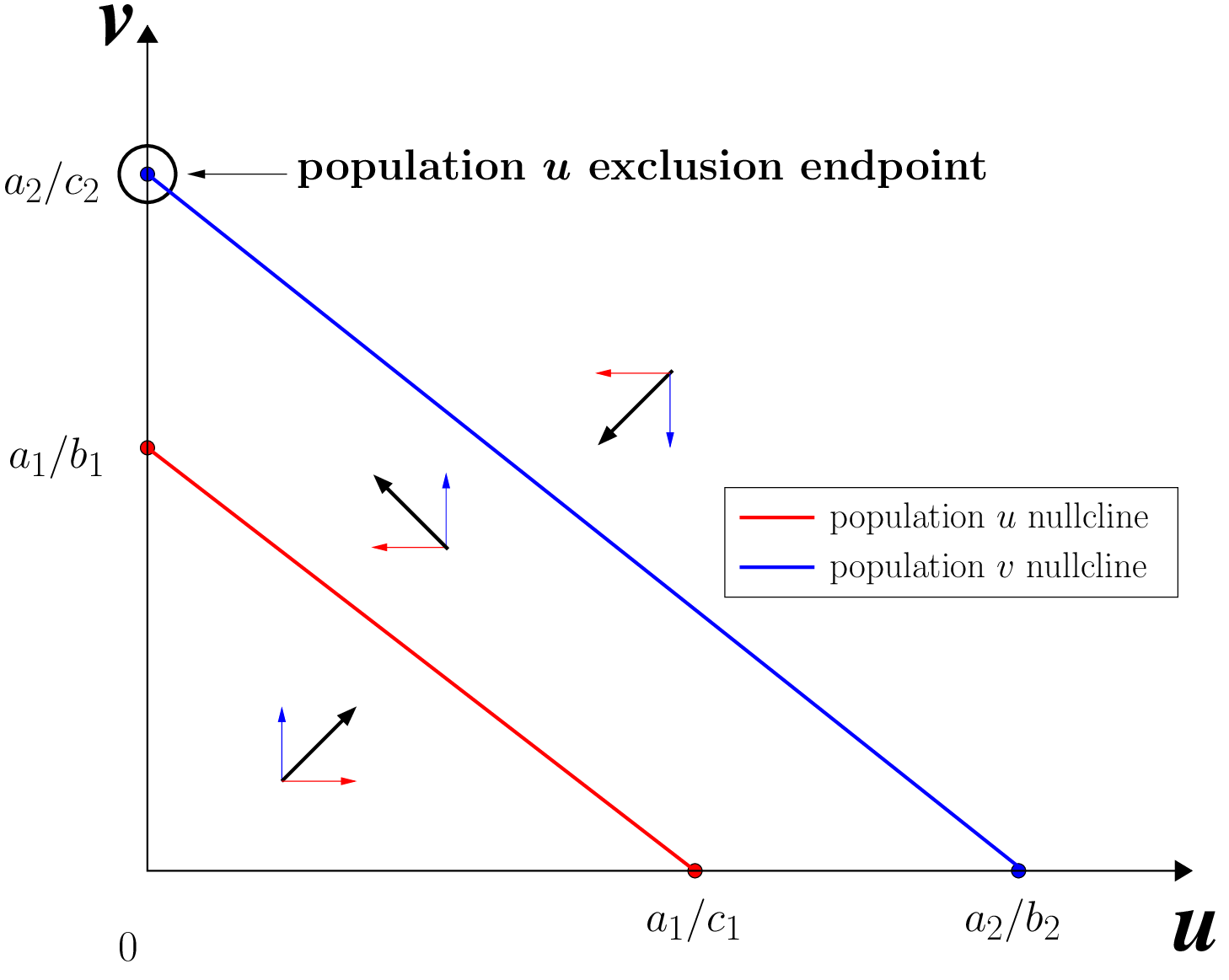} &
\includegraphics[scale=0.27]{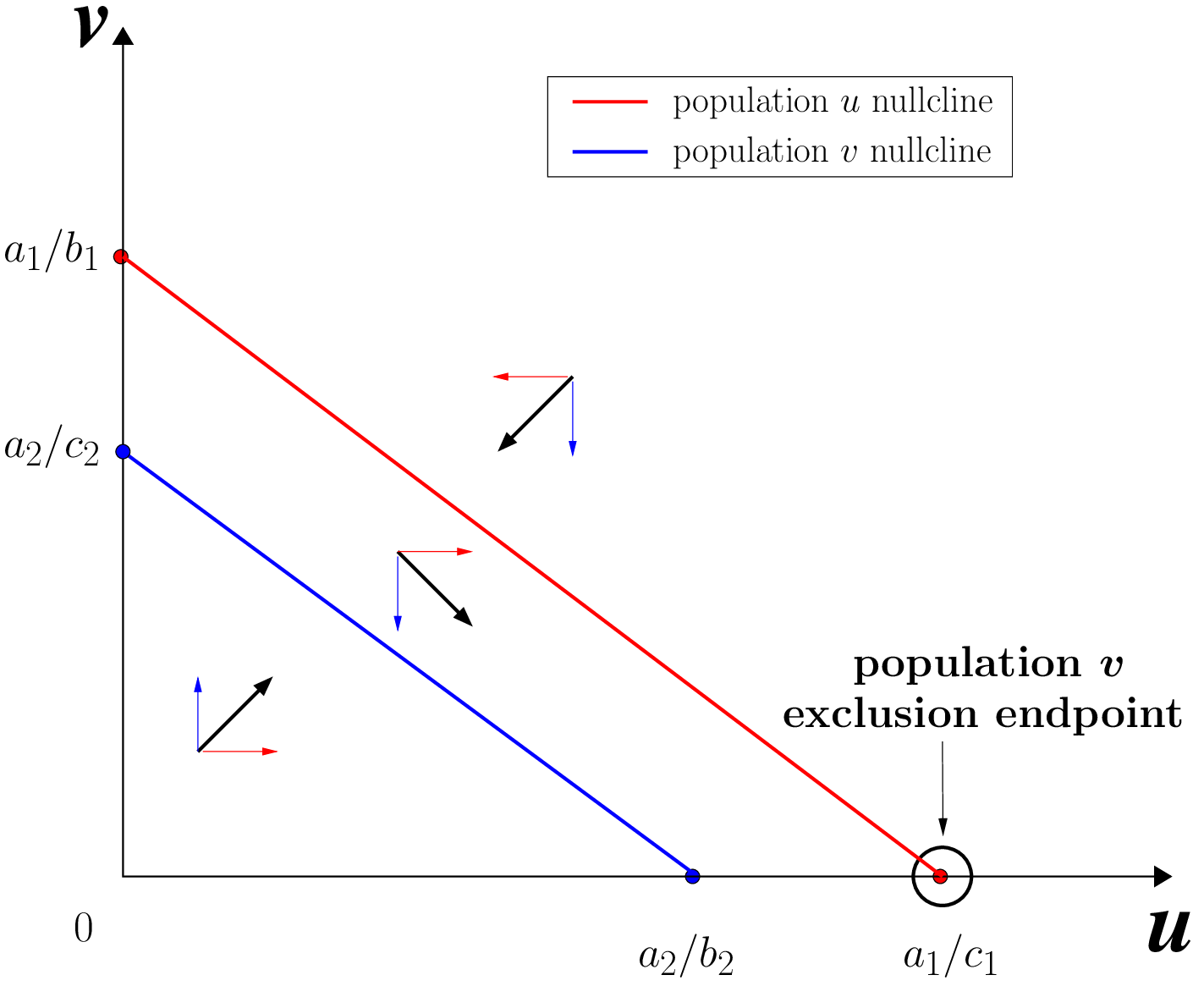} 
\end{array}$
\end{center}
\caption{\textit{Left}: Dynamic situations that leads to the exclusion
     of population $u$ (inequality \ref{exclusaou}). \textit{Right}:
     This situation means the elimination of the population $v$
     (inequality \ref{exclusaov}). In both cases the arrows colors mean
     the same as in Fig.\ \ref{stable-unstable}.}
\lb{exclusions}
\end{figure}

Having reviewed the basics of the Gause-Witt system dynamics we are
now in position to carry out an analysis of income distribution by
means of this model.

\section{Applying the Gause-Witt system to Brazil's income data}\lb{br}

It is a statement of the obvious that to survive all human populations
work for a living, either as hunters and gatherers collecting food and
other materials, or in farms, offices and industrial facilities. This
means that social groups can be viewed as sharing the same ecological
niche or ecosystem for survival. In the case of modern nation states
one may empirically assume that the ecosystem in which any modern
population lives actually forms an \textit{economic niche} or
\textit{economic ecosystem}. Under this viewpoint one may also consider
that the entire populations of modern nation states, formed by both rich
and poor individuals, share and consume the resources of the economic
niches they live in.

Bearing these thoughts in mind one may assume that the two sides of
the economic divide mentioned above, the 99\% and the 1\%, to be the
analogous of two ``species'' who live in the same economic ecosystem,
consume their resources and struggle for survival in a manner similar
to Gause's original study. Clearly this kind of analogy can be applied
to other economic groupings rather than the 99\% and 1\%, as well as to
other economic settings. In fact, in his seminal paper announcing the
competitive exclusion principle the American ecologist Garrett James
Hardin (1915-2003) mentioned some examples of possible applications of
this principle to economics \cite{hardin}. Here, however, we shall
remain focused on the 99\% and 1\% economic ``species'' with the aim of
testing the hypothesis of their \textit{economic struggle for existence}.
To do so we need to gather economic data which can then be attributed
to a competitive economic environment where an admittedly economic
struggle for existence takes place. Here we shall do so using Brazil's
income data already available in the literature from previous studies.

Moura Jr.\ \& Ribeiro \cite{nm09} studied the Brazilian income data
from 1978 to 2005 and concluded that $\approx$ 99\% of the individuals
sampled have their income fitted to the Gompertz curve, whereas the
remaining $\approx$ 1\% is well adjusted to the Pareto power-law
\cite[][tabs.\ 2 -- 3]{nm09}. These results were extended up to 2009
in Refs.\ \cite{fnm2010,nm2013}, which also provided figures regarding
the respective percentage shares of both the Gompertz and Pareto
regions to the total income sample.

The next step is to identify the income variables with those that can
be applied to the Gause-Witt system discussed above. Let us then
call $u\dn{\,99}$ the \textit{percentage share of the Gompertz region
(the 99\%)} with respect to the total income, and by $v\dn{1}$ the
\textit{percentage share of the Pareto power-law segment (the 1\%)}
also to the total income. Hence, $u\dn{\,99}$ and $v\dn{1}$ will be
the population sizes given in percentages of the two economic
``species'' that are being modeled here in a yet-to-be-verified
situation of economic competitive exclusion that should fall into one of
the four situations describe in Figs.\ \ref{stable-unstable} and
\ref{exclusions} and are summarized by the expressions (\ref{stable})
to (\ref{exclusaov}). The respective time derivatives $\dot{u}\dn{\,99}$
and $\dot{v}\dn{1}$ will also be needed. They can be numerically
calculated using the expression below,
\be
\dot{y} \approx \frac{y\,(t + \Delta t) - y\,(t - \Delta t)}{2\Delta t},
\lb{dfnum}
\ee
where $\Delta t$ stands for one year in the present modeling context.

Tab.\ \ref{tab1} presents Brazilian income data relevant to this paper
as collected from Refs.\ \cite{nm09,fnm2010,nm2013} in the period from
1981 to 2009. Because the numerical derivatives require data before and
after a certain year, the effective time span for the hypothesis testing
aimed here is from 1982 to 2008.
\begin{table*}[htbp]
\caption{Brazilian income data from 1981 to 2009 collected from Refs.\
	\cite{nm09,fnm2010,nm2013} with added time derivatives obtained
	with Eq.\ (\ref{dfnum}). Values in this table were evaluated
	directly from the raw data. The quantities from left to right
	stand for the sampling year, Gini coefficient, percentage
	share $u\dn{\,99}$ of the 99\% to total income (Gompertzian
	component), time derivative $\dot{u}\dn{\,99}$, percentage share
	$v\dn{1}$ of the 1\% to total income (Paretian component),
	and time derivative $\dot{v}\dn{1}$. Since there were no income
	samplings in 1991, 1994 and 2000 \cite[see][]{nm09} some results
	for these years were obtained by numerical interpolation.
        Dimensions are given and uncertainties were not provided due to
        data gathering limitations.\lb{tab1}}
\begin{center}
\footnotesize
\begin{tabular}{ccccccccccc}
\hline\noalign{\smallskip}
year & $Gini$ & $u\dn{\,99} \, (\%)$ & $\dot{u}\dn{\,99} \, 
(\%/\mbox{year})$ & $v\dn{1} \, (\%)$ & $\dot{v}\dn{1} \, (\%/\mbox{year})$
\\ 
\noalign{\smallskip}\hline\noalign{\smallskip}
1981&$0.574$&$87,662$ &    ---    &$12.338$ &     ---   \\
1982&$0.581$&$87,150$ & $+1.0815$ &$12.850$ & $-1.0815$ \\
1983&$0.584$&$85,499$ & $-0.0420$ &$14.501$ & $+0.0420$ \\
1984&$0.576$&$87,234$ & $-0.1740$ &$12.766$ & $+0.1740$ \\
1985&$0.589$&$85,847$ & $+0.9945$ &$14.153$ & $-0.9945$ \\
1986&$0.580$&$85,245$ & $-0.0490$ &$14.755$ & $+0.0490$ \\
1987&$0.592$&$85,945$ & $-0.0785$ &$14.055$ & $+0.0785$ \\
1988&$0.609$&$85,402$ & $+1,7355$ &$14.598$ & $-1.7355$ \\
1989&$0.628$&$82,474$ & $-0.2340$ &$17.526$ & $+0.2340$ \\
1990&$0.605$&$85,870$ & $-1.9833$ &$14.130$ & $+1.9833$ \\
1991&  ---  &$86.441$ & $-0.5705$ &$13.560$ & $+0.5705$ \\
1992&$0.578$&$87.011$ & $+1.1843$ &$12.989$ & $-1.1843$ \\
1993&$0.599$&$84.072$ & $+1.0083$ &$15.928$ & $-1.0083$ \\
1994&  ---  &$84.994$ & $-0.9225$ &$15.006$ & $+0.9225$ \\
1995&$0.596$&$85.917$ & $-0.8568$ &$14.083$ & $+0.8568$ \\
1996&$0.598$&$86.708$ & $-0.1160$ &$13.292$ & $+0.1160$ \\
1997&$0.598$&$86.149$ & $+1.0935$ &$13.851$ & $-1.0935$ \\
1998&$0.597$&$84.521$ & $+0.0835$ &$15.479$ & $-0.0835$ \\
1999&$0.590$&$85.982$ & $-0.5323$ &$14.018$ & $+0.5323$ \\
2000&  ---  &$85.586$ & $+0.3965$ &$14.414$ & $-0.3965$ \\
2001&$0.592$&$85.189$ & $-0.4058$ &$14.811$ & $+0.4058$ \\
2002&$0.586$&$86.397$ & $-0.0820$ &$13.603$ & $+0.0820$ \\
2003&$0.579$&$85.353$ & $-0.3985$ &$14.647$ & $+0.3985$ \\
2004&$0.582$&$87.194$ & $-0.4370$ &$12.806$ & $+0.4370$ \\
2005&$0.580$&$86.227$ & $-0.2590$ &$13.773$ & $+0.2590$ \\
2006&$0.592$&$87.712$ & $+0.2685$ &$12.288$ & $-0.2685$ \\
2007&$0.572$&$85.690$ & $+0.2790$ &$14.310$ & $-0.2790$ \\
2008&$0.543$&$87.154$ & $-0.3575$ &$12.846$ & $+0.3575$ \\
2009&$0.539$&$86.405$ &   ---     &$13.595$ &     ---   \\
\noalign{\smallskip}\hline
\end{tabular}
\end{center}
\end{table*}
Now, in order to fit the data shown in Tab.\ \ref{tab1} to Eqs.\
(\ref{f1gw}) and (\ref{f2gw}) let us first rewrite them as follows,
\bse
\begin{empheq}[left=\empheqlbrace]{align}
\displaystyle\frac{\dot{u}\dn{\,99}}{u\dn{\,99}}&=
	A_1+B_1\,v\dn{1}+C_1\,u\dn{\,99}, \lb{f1gw991} \\[-8pt] \nonumber \\
\displaystyle\frac{\dot{v}\dn{1}}{v\dn{1}}&=
	A_2+B_2\,u\dn{\,99}+C_2\,v\dn{1}. \lb{f2gw991}
\end{empheq}
\ese
Fitting these expressions to the data in Tab.\ \ref{tab1} is a
straightforward case of multiple regression analysis. The fitting was
carried out in three time spans:
\begin{enumerate}[(I)]
	\item 1982 to 2008, \qquad (whole time span);
	\item 1982 to 1994, \qquad (1st subperiod);
	\item 1995 to 2008, \qquad (2nd subperiod).
\end{enumerate}

The reasoning underlying the consideration of two subperiods lies
in the fact that Brazil experienced runaway inflation and hyperinflation
for over two decades, which came to an abrupt end in 1994. Fig.\ 2 of
Ref.\ \cite{nm2013} shows the phase portrait of the two variables of
the Goodwin system for Brazil where one can notice that their cycling
pattern actually moved from the lower region to the higher one on the
right exactly after 1994. This seems to indicate that the end of
hyperinflation produced a shock in the economic system, a fact that
suggests the present analysis in two subperiods.

\subsection{Case I}

Eqs.\ (\ref{f1gw991})-(\ref{f2gw991}) fitted to the data in Tab.\
\ref{tab1} from 1982 to 2008 are as follows.

\begingroup
\footnotesize
\be
\left\{
\begin{aligned}\lb{caso1}
A_1&=-0.18\pm0.26, \\
B_1&=0.0011\pm0.0014,\\ 
C_1&=0.00086\pm0.00183,\\ 
R^2&=0.024, 
\end{aligned}
\right.
\text{\qquad and \qquad}
\left\{
\begin{aligned}
A_2&=7.130\pm0.092, \\
B_2&=-0.000439\pm0.000566,\\ 
C_2&=-0.0709\pm0.0008,\\ 
R^2&=0.997, 
\end{aligned}
\right.
\ee
\endgroup
Comparing the parameters of Eqs.\ (\ref{f1gw})-(\ref{f2gw}) with those
in Eqs.\ (\ref{f1gw991})-(\ref{f2gw991}) and considering the signs we
may conclude that this case yields,
\be
\begin{aligned}\lb{rescaso1}
\frac{a_1}{b_1} > \frac{a_2}{c_2}, 
\end{aligned}
\text{\qquad and \qquad}
\begin{aligned}
\frac{a_1}{c_1} < \frac{a_2}{b_2}. 
\end{aligned}
\ee
This corresponds to the dynamic state of \textit{stable coexistence}
indicated in the left Fig.\ \ref{stable-unstable}.

\subsection{Case II}

Eqs.\ (\ref{f1gw991})-(\ref{f2gw991}) fitted to the data in Tab.\
\ref{tab1} from 1982 to 1994 are as follows:

\begingroup
\footnotesize
\be
\left\{
\begin{aligned}\lb{caso2}
A_1&=-0.31\pm0.44, \\
B_1&=0.0015\pm0.0023,\\ 
C_1&=0.0020\pm0.0033,\\ 
R^2&=0.048, 
\end{aligned}
\right.
\text{\qquad and \qquad}
\left\{
\begin{aligned}
A_2&=7.127\pm0.145, \\
B_2&=-0.000947\pm0.000988,\\ 
C_2&=-0.0705\pm0.0012,\\ 
R^2&=0.997, 
\end{aligned}
\right.
\ee
\endgroup
which yield,
\be
\begin{aligned}\lb{rescaso2}
\frac{a_1}{b_1} > \frac{a_2}{c_2}, 
\end{aligned}
\text{\qquad and \qquad}
\begin{aligned}
\frac{a_1}{c_1} < \frac{a_2}{b_2}.
\end{aligned}
\ee
These inequalities also correspond to the dynamic state of
\textit{stable coexistence}.

\subsection{Case III}

Eqs.\ (\ref{f1gw991})-(\ref{f2gw991}) fitted to the data in Tab.\
\ref{tab1} from 1995 to 2008 are as follows.

\begingroup
\footnotesize
\be
\left\{
\begin{aligned}\lb{caso3}
A_1&=-0.00470\pm0.279, \\
B_1&=5.237\pm0.002,\\ 
C_1&=-1.900\pm0.002,\\ 
R^2&=0.00007, 
\end{aligned}
\right.
\text{\qquad and \qquad}
\left\{
\begin{aligned}
A_2&=7.178\pm0.096, \\
B_2&=0.000259\pm0.000488,\\ 
C_2&=-0.0720\pm0.0008,\\ 
R^2&=0.998, 
\end{aligned}
\right.
\ee
\endgroup
which yield,
\be
\begin{aligned}\lb{rescaso3}
\frac{a_1}{b_1} < \frac{a_2}{c_2}, 
\end{aligned}
\text{\qquad and \qquad}
\begin{aligned}
\frac{a_1}{c_1} > \frac{a_2}{b_2}.
\end{aligned}
\ee
These inequalities correspond to the \textit{unstable coexistence}
dynamic according to the right Fig.\ \ref{stable-unstable}.
Nevertheless, this case ought to be rejected as unrealistic because the
right inequality in the expressions (\ref{rescaso3}) produced
negative values.

\section{Discussion}\lb{discu}

The results presented above clearly show a goodness of fit discrepancy
in all three cases above as measured by the coefficient of determination
$R^2$. Eq.\ (\ref{f1gw991}) performed poorly against the data while Eq.\
(\ref{f2gw991}) performed well according to this coefficient. Since these
two ODEs are coupled, such discrepancy suggests problems with the data
rather than an indication of the null hypothesis for Eq.\ (\ref{f1gw991}).
A possible explanation would be that the analysis carried out here used
legacy data from Refs.\ \cite{nm09,fnm2010,nm2013} whose aims were
different from the testing proposed here. So, striving for better data
for further testing is the subject of future work.

Secondly, Cases I and II indicate stable coexistence, an expected result
in a politically stable capitalist country like Brazil, which at present,
and for quite some time, does not have any major internal political
upheaval, does not face problems with its neighboring countries, and is
not involved in any external conflict. So, considering this context it
would be surprising if the results were to reveal something different
than stable coexistence between the 99\% and the 1\% in Brazil. Then, a
possible explanation for the unstable coexistence arising in Case III
and its dismissal after producing unrealistic results could be the
economic shock produced by the abrupt end of hyperinflation. Maybe the
time series after 1994 is not long enough to reach the point of aftershock
relaxation time. Better and more up to date data should clarify this point.

Regarding the competitive exclusion principle discussed in Sec.\
\ref{gw}, one notices that when Gause's statement (1) is applied to
the 99-1\% economic divide the stable coexistence actually means that
these two ``species'' must be doing things differently in economic
terms. In other words, they must be utilizing differently the economic
resources of their shared economic niche and their growth is somehow
limited. In fact, this seems what really happens, because it has
already been pointed out by various authors that taxation of the 1\%
is essential to avoid the rich ending up with all resources \cite{piketty}
\cite[Secs.\ 4.2.6, 4.2.7, and references therein]{ribeiro2020income},
which in the context of this study would mean taking the 99\% to the
exclusion point, that is, extinction, as shown in the left Fig.\
\ref{exclusions}.

Another point regarding the interpretation of Gause's principle in the
context of economics and econophysics is that the 1\% does not live on
wages, at least not on ordinary wage values typically paid to the 99\%,
but on investment \cite[Secs.\ 4.3, 4.4]{ribeiro2020income}
\cite{scaf02,scaf04a,scaf04b}, extremely high salaries obtained by top
actors and senior managers of top firms, and from capital return often
accumulated by previous generations \cite{piketty} \cite[Chap.\ 3]
{ribeiro2020income}. So, these two economic ``species'' are in fact
doing things differently in economic terms, and because of that they
are able to live in stable coexistence according to Gause's principle.
If one of these two ``species'' moves too much onto the other's
``grounds,'' so to speak, the system runs the risk of breakdown, which
in the present context means that one of the two final situations
depicted in Fig.\ \ref{exclusions} ought to take place.

Finally, the parameters presented in the expressions of the Gause-Witt
model are simplified versions of the ones appearing in ecological
studies. This is so because in population ecology those parameters
are in fact combinations of others representing population growth,
parasitism, carrying capacity, population density, etc. The Goodwin
model started with several economic quantities whose ingenious
combination led to a predator-prey ODE system. This has not been done
here due to the different route taken in this investigation.

As mentioned above, the work of Ref.\ \cite{nm2013} took the opposite
path than followed by Goodwin \cite{g67}, in the sense that Ref.\
\cite{nm2013} started with the ODE system and then applied the data to
the model. As this is the same investigative route taken here, the
economic interpretation of the Gause-Witt parameters, actually their
most likely unfolding into other quantities, possibly requires the
opposite investigative path, that is, to start with variables such as
tax rates, wage levels and investment returns, and then somehow work
out how these quantities can be related among themselves such that
they produce the Gause-Witt ODE system. At present there is no obvious
way on how to do that and, therefore, the economic interpretation of
the Gause-Witt parameters is an open problem.

\section{Conclusions}\lb{conc}

This work explored the viability of the Gause-Witt model of population
ecology and ecosystems to represent the competitive income distribution
dynamics of social classes in economic systems. The basic feature of
this model, which is a coupled nonlinear first-order ordinary
differential equations that represents species sharing the same
ecological niche and competing for the same resources, was applied to
the income data of Brazil in order to test the dynamic of the two well
defined segments of the population as far as income data is concerned:
the 99\% and the 1\%. These two income segments arouse from previous 
analyses \cite{nm09,fnm2010,nm2013} of the complementary cumulative
distribution functions (CCDFs) built upon yearly datasets that showed
two distinct income regions: the lower income comprising about 99\%
of the population, which in Brazil's case is better represented by the
Gompertz curve, and the richest 1\% described by the Pareto power-law
at the CCDF's tail. 

After presenting the basic dynamic features of the Gause-Witt system
and discussing its concepts in the context of population ecology
dynamics, the model was applied to the income data of Brazil using
legacy datasets provided by previous studies \cite{nm09,fnm2010,nm2013}
in the period from 1981 to 2009. Multiple regression analyses were
carried out with the data in three periods: (I) 1982 to 2008, (II) 1982
to 1994 and (III) 1995 to 2008.

The results arising from cases I and II indicate that the 99\% and
1\% income classes are mostly in the dynamic state of stable
coexistence as far as Brazilian data is concerned in the studied
time period. The data fits for the subperiod 1995--2008 resulted in
unrealistic parameters and was then dismissed. The goodness of fit
for the cases I and II showed discrepancies between the two equations
of the Gause-Witt system, a situation which indicates that better data
are needed to reach at more conclusive evidence regarding the shared
dynamic state of these two economic segments.

Despite the limitations of the results portrayed here it seems that
the Gause-Witt system presents good prospects in terms of modeling
economic competition, prospects which are worth further investigations
not only with better Brazilian data and at longer time spans, but also
at other countries and regions. In fact, the model can in principle be
applied to other settings rather than the 99-1\% divide, that is, to
other economic situations where one can model competitive dynamics
among players sharing the same economic ecosystem.

In the context of the present paper it should also be mentioned other
recent works that consider in different ways the terribly skewed
income and wealth distributions. On this respect Ghosh and
collaborators \cite{ghosh2023} showed numerical simulations of
kinetic exchange models that lead to extreme inequalities, where
about 13\% of the population possesses about 87\% of the total wealth,
a result in line with recently reported data \cite{oxfam3}. Similar
situation was discussed by Greenberg \& Gao \cite{greenberg2024},
who showed that wealth is much more unequally distributed than income,
a conclusion also reached by Piketty \citetext{\citealp{piketty};
see also \citealp{ribeiro2020income}, Chap.\ 3}. In addition, in
reviewing the extensive econophysics literature on random asset
exchange models Greenberg \& Gao \cite[p.\ 18]{greenberg2024} stated
that ``a large proportion of observed economic inequality is the
result of luck and the inherently diffusive (entropy-increasing)
nature of exchange itself, and not the result of interpersonal
differences in industriousness, entrepreneurialism, or intelligence'',
concluding that ``large scale regimes of wealth redistribution, such
as wealth taxes, may be necessary in order to reduce inequality'', a
viewpoint on inequality reduction similar to Piketty's \cite{piketty}.
Hence, the present study adds to the growing literature which indicates
the acuteness of the present inequality scenario.

As final comments, this paper put forward a model that adapts concepts
and equations developed in a certain area to another, in this case from
population ecology to economics and econophysics. Conceptual transportation
among different research areas can be very fruitful, but can also be
fraught with misinterpretations. For instance, the Goodwin model was
initially thought to be a ``class struggle model'' because its predator-prey
variables were viewed as representing a ``struggle'' between workers
and capitalists. However, further analyses showed that its variables
can also be interpreted differently, as a predator-prey system between
employed and unemployed workers, and still as quantities of capital and
consumption goods \cite[][Secs.\ 5.3.3, 6.5.1, and references therein]
{ribeiro2020income}. Hence, bearing in mind possible caveats, one may
argue that the 99-1\% social classes divide form a symbiotic
relationship which in view of the present study could be classified as
mutualistic and, perhaps, also somewhat commensalistic or parasitic.

\section*{Dedication}

To the memory of my colleague and collaborator Newton Jose de Moura Jr.\
(1969-2021), who shared the excitement for the research in econophysics.

\section*{Acknowledgments}

Partial financial support was received from FAPERJ -- \textit{Rio de Janeiro State 
Research Funding Agency}, grant number E-26/210.552/2024.

\section*{Competing Interests}

The author declares that he has no known competing financial interests
or personal relationships that could have appeared to influence the
work reported in this paper.

\section*{Data Availability Statement}

This paper has no associated data.









\bibliography{econphys}
\bibliographystyle{elsarticle-num}
\end{document}